\documentclass[prd,nofootinbib,showpacs,showkeys,twocolumn,a4paper]{revtex4}

\pdfoutput=1

\usepackage{graphicx}
\usepackage{amsmath,amssymb}
\usepackage{amsfonts}
\usepackage{latexsym}
\usepackage{mathrsfs}
\usepackage{color}
\usepackage{slashed}
\usepackage{float}
\usepackage{feyn}
\usepackage[all]{xy}
\usepackage{feyn}
\usepackage{hyperref}
\usepackage{dsfont}
\usepackage{placeins}
\usepackage{cancel}

\allowdisplaybreaks

\definecolor{blue}{rgb}{0,0,0.5}
\definecolor{lightblue}{rgb}{0,0,1}
\definecolor{red}{rgb}{0.5,0,0}
\definecolor{lightred}{rgb}{1,0.5,0}
\definecolor{green}{rgb}{0,0.5,0}
\definecolor{darkgreen}{rgb}{0.0,0.3,0.0}
\definecolor{orange}{rgb}{1,0.4,0}
\definecolor{grey}{rgb}{0.5,0.5,0.5}
 
\providecommand{\eqn}{Eq.~}

\providecommand{\fig}{Fig.~}

\providecommand{\sect}{Sec.~}

\newcommand{\eprint}[1]{\href{http://arxiv.org/abs/#1}{#1}}

\newcommand{\printifnonempty}[2]{\if\relax\detokenize{#1}\relax\else#2\fi}
\newcommand{\alter}[5]{%
  \long\def\temp{#3}%
  \long\def\accept{#5}%
  \ifx\temp\accept
    {#1}
  \else
    {\textcolor{#4}{\printifnonempty{#1}{{#1}}}%
    \textcolor{grey}{\printifnonempty{#2}{(#2)}}%
    \textcolor{#4}{\printifnonempty{#3}{{[#3]}}}}%
  \fi
}

\global\long\def\m{\mu}
\global\long\def\e{\epsilon}
\global\long\def\n{\nu}
\global\long\def\d{\partial}

\global\long\def\l{\lambda}

\global\long\def\a{\alpha}

\global\long\def\b{\beta}
\global\long\def\a{\alpha}

\global\long\def\b{\beta}

\newcommand{\ev}[1]{\ensuremath{\left\langle #1 %
                     \right\rangle}} 

\begin{document}



\title{Conformal Inverse Seesaw and Warm Dark Matter}

\title{The Inverse Seesaw in Conformal Electro-Weak Symmetry Breaking\\ and Phenomenological Consequences}

\author{Pascal Humbert}
\email[\,]{humbert@mpi-hd.mpg.de} 

\author{Manfred Lindner}
\email[\,]{manfred.lindner@mpi-hd.mpg.de}   

\author{Juri Smirnov}
\email[\,]{juri.smirnov@mpi-hd.mpg.de}

\affiliation{
Max-Planck-Institut f\"ur Kernphysik, Saupfercheckweg 1, 69117 Heidelberg, Germany}

\begin{abstract}

\noindent
We study the inverse seesaw mechanism for neutrino masses and phenomenological consequences in the context of conformal electro-weak symmetry breaking. The main difference to the usual case is that all explicit fermion mass terms including Majorana masses for neutrinos are forbidden. All fermion mass terms arise therefore from vacuum expectation values of suitable scalars times some Yukawa couplings. This leads to interesting consequences for model building, neutrino mass phenomenology and the Dark Matter abundance. In the context of the inverse seesaw we find a favoured scenario with heavy pseudo-Dirac sterile neutrinos at the TeV scale, which in the conformal framework conspire with the electro-weak scale to generate keV scale warm Dark Matter. The mass scale relations provide naturally the correct relic abundance due to a freeze-in mechanism. We demonstrate also how conformal symmetry decouples the right-handed neutrino mass scale and effective lepton number violation. We find that lepton flavour violating processes can be well within the reach of modern experiments. Furthermore, interesting decay signatures are expected at the LHC.

\end{abstract}

\maketitle

\section{\label{sec:Introduction}Introduction}

\noindent
So far there are no signs for new physics beyond the Standard Model (SM) at the LHC and this 
raises the question if mechanisms of electro-weak symmetry breaking are at work which are different 
from what was expected over the last decades. The SM has besides the vacuum expectation value 
no explicit scale and this, as well as the fact that the quartic Higgs coupling runs very close to zero at 
the Planck-scale, may point to a role of conformal symmetry and its breaking by quantum effects.
No explicit mass scales would therefore be allowed in the Lagrangian and symmetry breaking would 
be the consequence of a Coleman Weinberg type mass generation \cite{Coleman:1973jx}. This is from 
a technical point of view more restrictive since the number of allowed terms in the Lagrangian is reduced. 
The minimal phenomenological scenarios require therefore some extra fields and various specific models 
have been worked out \cite{Coleman:1973jx,Fatelo:1994qf,Hempfling:1996ht,Hambye:1995fr,Meissner:2006zh,Foot:2007as,Foot:2007ay,Chang:2007ki, Hambye:2007vf, Meissner:2007xv, Meissner:2009gs,Iso:2009ss,Holthausen:2009uc, Iso:2009nw,Foot:2010et, Khoze:2013uia,Kawamura:2013kua,Gretsch:2013ooa, Heikinheimo:2013fta,Gabrielli:2013hma,Carone:2013wla,Khoze:2013oga,Englert:2013gz,Farzinnia:2013pga,Abel:2013mya, Foot:2013hna,
Hill:2014mqa,Guo:2014bha,AlexanderNunneley:2010nw,Radovcic:2014rea,Khoze:2014xha,Smirnov:2014zga,Salvio:2014soa, Kannike:2014mia,Kannike:2015apa,Chankowski:2014fva,Okada:2014nea,Guo:2015lxa,Baek:2015mna,
Hatanaka:2014tba,Kang:2014cia,Cai:2014kra,Benic:2014aga,Gorsky:2014una,Okada:2014qsa,Okada:2014oda,Khoze:2014woa,Lattanzi:2014mia}.  

In this paper we study consequences for the neutrino sector arising from the fact that explicit fermion 
mass terms (both Dirac and especially also Majorana) are no longer allowed when the SM is extended 
to incorporate neutrino masses and mixings. All Dirac and Majorana mass terms must then stem from 
Yukawa couplings times vacuum expectation values of suitable scalars. Specifically we present a simple 
extension of the SM which can account for non-zero neutrino masses and which leads to spontaneous 
conformal symmetry breaking. At the same time the extension of the scalar sector is such that no low-scale 
Landau pole appears. We will show that our set-up leads naturally to the inverse seesaw (ISS) scenario, 
with active-sterile mixing at a phenomenologically interesting level, where over-all electro-weak fits are
improved \cite{Akhmedov:2013hec,Akhmedov:2016xna,Dev:2014xea,Abada:2014nwa}.
Furthermore, we will find that the UV completion of the theory, in particular the requirement of anomaly cancellation, forces us to introduce additional fermions, which turn out to have a lifetime, which makes them potential Dark Matter (DM) candidates. 

The paper is organized in the following way. In \sect \ref{sec:Generics} we will discuss the generic features of conformal model building and implications for Dark Matter and neutrino mass phenomenology. In \sect \ref{sec:Model} we will present a concrete UV complete and anomaly free extension of the Standard Model, which we consider as well motivated by precision electro-weak data. Using this model as an example we will demonstrate in \sect \ref{sec:RCSB}  how in particular the Radiative Conformal Symmetry Breaking (RCSB) works in this set-up and what it implies for the particle spectrum of the model. In \sect \ref{sec:Phenomenology} we discuss the low energy particle phenomenology of the model and several Dark Matter production mechanisms.   We will summarize our findings in \sect \ref{sec:Conclusion}.
  
\section{\label{sec:Generics} Generics of Conformal Model Building}
\noindent
The idea of spontaneous conformal symmetry breaking is rather old and was put forward by Coleman and Weinberg \cite{Coleman:1973jx}. It has been argued by Bardeen \cite{Bardeen:1995kv}, that its protective feature can avoid the fine-tuning due to quadratic divergences and thus keep the Higgs mass safe, as it is only multiplicatively renormalized. The same argument applies in curved space-time background to diffeomorphism symmetry \cite{Smirnov:2014zga} and can protect the vacuum energy from power divergences. For this mechanism to work, however, the particle content of the theory needs to be specified in order to explain the RG running necessary for the RCSB.

It is, for example, clear that this mechanism can not be at work in the standard model for a top mass above 79~GeV, as the fermionic contribution drives the potential couplings in a way which does not allow for RCSB. It is therefore clear that an extension of the SM by some Hidden Sector (HS) is necessary. The HS can be coupled to the SM via different portals: One portal is connected to neutrino masses. We know that neutrino masses are finite and SM singlet fermions which can connect to a HS are therefore well motivated. Further portal operators arise from the $H^\dagger H $ singlet combination of the Higgs field with other scalars $H'$ by renormalizable $H^\dagger H H'^\dagger H' $ quartic interactions or the kinetic mixing of the photon with an additional massive $U(1)$ gauge boson. 

As we will see, the existence of the Higgs portal is an absolute necessity for any RCSB model to work.
Another crucial requirement for the HS is the mass dominance of bosonic degrees of freedom in order to achieve RCSB due to the RG running.  
Furthermore, the couplings should be such that no Landau pole appears at an adjacent energy scale making the theory ill defined. Another important feature is to make sure that the quartic couplings of the potential remain positive from the high scale on throughout all the RG running. It is obvious that vacuum stability is a built-in feature of such a model.  

The HS itself can contain a Hidden Symmetry group which can be gauged. This additional structure may be used to explain the smallness of the active neutrino masses and we will demonstrate an example of this in the next section. The gauge symmetry needs to be anomaly free
which implies additional constraints on the particle spectrum and may lead to the existence of long-lived particles which can be Dark Matter candidates. In the next section we present a model, in which  active neutrinos acquire their mass in an inverse seesaw mechanism, which owes its mass matrix structure to the HS $U(1)$ symmetry. This symmetry is gauged and the anomaly freedom condition requires us to have a particle content which contains a long-lived particle. We observe that imposing constraints from low energy particle physics leads to a parameter region with a warm Dark Matter candidate compatible with all astrophysical observations. Furthermore, several production mechanisms can account for the correct relic density in our model.

\section{\label{sec:Model} The Conformal Inverse Seesaw}     
\noindent
We will demonstrate the features described above using an explicit model which was introduced in \cite{Lindner:2014oea}. The Conformal Inverse Seesaw (CISS) has the gauge group $SU(3)_c\times SU(2)_L \times U(1)_Y \times U(1)_X$. The scalar field content is extended by two SM singlet fields $\phi_1$ and $\phi_2$ with $U(1)_X$ charges one and two, respectively, and the potential \eqn \ref{eqn:potential}.  The fermion sector contains a total singlet field $\nu_R$ and a pair of chiral SM singlet fields $N_L$ and $N_R$, those, however, carry one unit of $U(1)_X$ charge. Note that the existence of the pair of fields with identical $U(1)_X$ charge is required by anomaly cancellation.
\begin{widetext}
\begin{align}
\label{eq:TheModel}
& \mathcal{L}_\text{CISS} = i \, \bar{N}_L \left( \slashed{\d} - i\,g_X \,X^\m \,\gamma_\m \right)\,N_L + i \, \bar{N}_R \left( \slashed{\d} - i\,g_X \,X^\m \,\gamma_\m \right)\,N_R - \frac{\tilde{y}_1}{2} \left( \bar{N}_R^c\,\nu_R \,\phi_1^* + h.c.\right)- \frac{y_1}{2} \left( \bar{N}_L\,\nu_R \,\phi_1 +h.c.\right)  \nonumber \\ 
& - \frac{y_2}{2} \left( \bar{N}_L\,N_L^c \,\phi_2 +h.c. \right) - \frac{\tilde{y}_2}{2} \left( \bar{N}_R\,N_R^c \,\phi_2 +h.c. \right)+ \frac{y_D}{2} \left( \bar{L}\,\tilde{H} \nu_R +h.c. \right)  \nonumber \\
&+ |\left( \d_\m -2 \,i\,g_X\,X_\m \right) \phi_2|^2 + |\left( \d_\m - \,i\,g_X\,X_\m \right) \phi_1|^2 -\frac{1}{4} F_X^{\m\n}F^X_{\m\n} + \frac{\kappa}{4} F_X^{\m\n}\,F_{\m\n} - V \left( H, \phi_1, \phi_2 \right)\,.
\end{align}
\end{widetext}
 We furthermore assume a  L-R exchange symmetry in the Hidden Sector i.e$.$ $N_L \leftrightarrow N_R^c$ which fixes the relations among the Yukawa couplings $y_1 = \tilde{y}_1$ and $y_2 = \tilde{y}_2$.

\begin{table}[h]
\begin{tabular}{c|c|c|c|c|c|c|c}
\hline 
 & $H$ & $\phi_1$ & $\phi_2$ & $L$ &  $\nu_R$ & $N_R$ & $N_L$\\
\hline 
\hline 
$U(1)_X$ & 0 & 1 & 2 & 0 & 0 & 1 & 1\\
Lepton Number & 0 & 0 & 0 & 1 & 1 & 0 & 0 \\
$U(1)_{Y}$ & 1 & 0 & 0 & -1 & 0 &0 & 0 \\
$SU(2)_{L}$ & 2 & 1 & 1 & 2  & 1 & 1 & 1 \\
\hline 
\end{tabular}
\caption{\label{tab:QuantumNumbers}
Quantum numbers in the Conformal Inverse Seesaw.}
\end{table}

\begin{figure}[h]
  \includegraphics[width=0.45\textwidth]{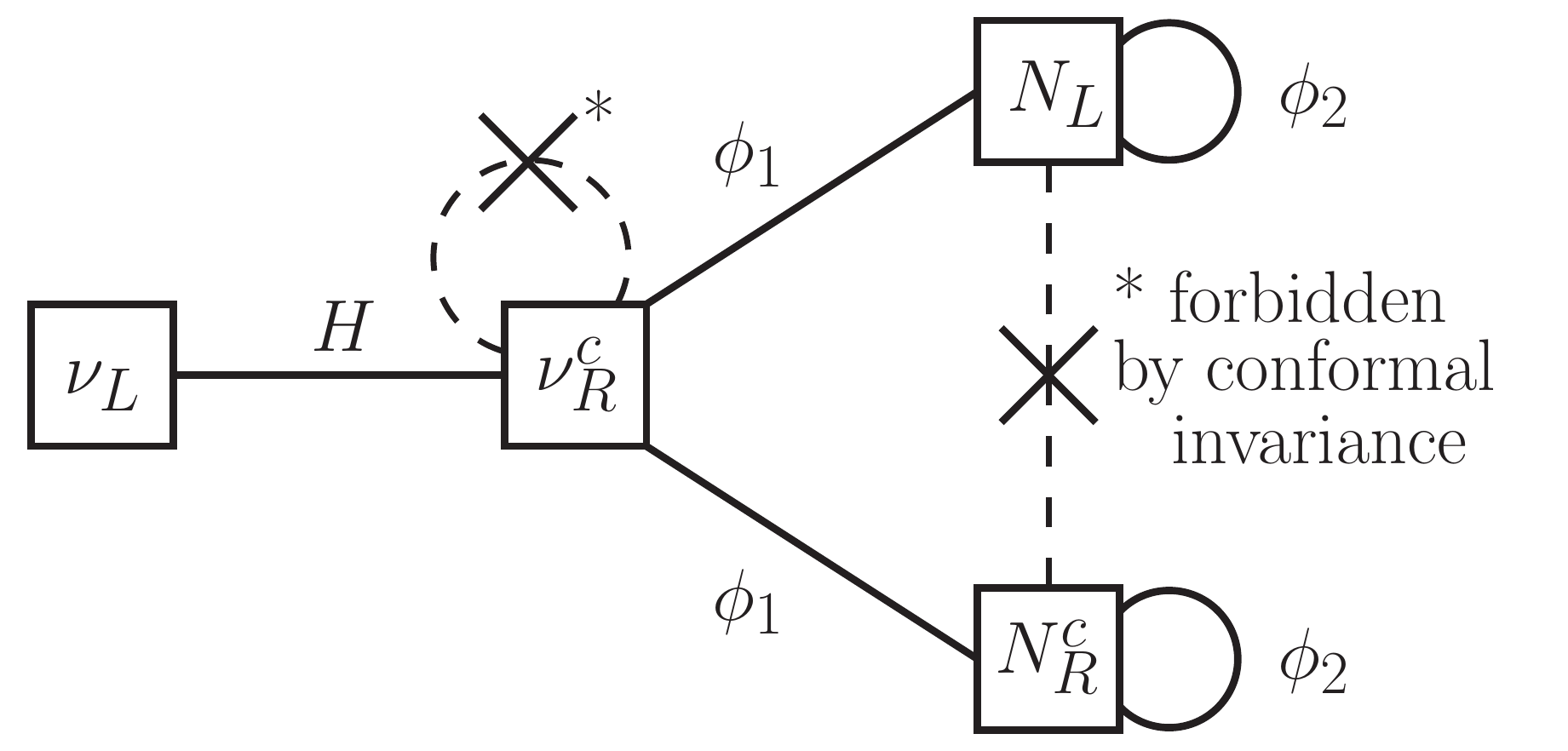}
  \caption{The diagrammatic visualisation of mass relations in the CISS. Conformal invariance forbids masses for singlet field combinations, which would be allowed in any non conformal theory with the same scalar spectrum.} 
  \label{fig:CISS_Masses}
\end{figure}

The scalar potential contains all combinations allowed by the quantum numbers

\begin{align}
\label{eqn:potential}
V(H, \phi_1, \phi_2) = \frac{\l_H}{2}\,(H^\dagger\,H)^2 + \frac{\l_1}{2}\,\phi_1^4 + \frac{\l_2}{2} \,\phi_2^4 + \\ \nonumber      \l_{H1}\,H^\dagger\,H\,\phi_1^2 + \l_{H2}\,H^\dagger\,H\,\phi_2^2 +\l_{1\,2}\,\phi_2^2\,\phi_1^2\,.
\end{align}
As we will elaborate on shortly, radiative effects break the conformal symmetry and all scalars acquire vacuum expectation values (vevs). This causes the breaking of $SU(2)_L\times U(1)_Y$ symmetry and leads to massive electro-weak gauge bosons. At the same time the breaking of $U(1)_X$ generates a mass for the Hidden Sector gauge boson.
 Using the compact notation in the basis $N^T = \left(  \nu_L , \nu_R^c, N_L, N_R^c \right)$ the mass term of the form $1/2\,\sum_{ij}\,\mathcal{M}_{ij} \,\bar{N}_i N_j^c$ has the following mass matrix

\begin{equation}
 \label{eq:seesawInvDM}
        \mathcal{M} = 
        \begin{pmatrix}
          0 & y_D \, \ev{H} & 0 & 0 \\
          y_D \, \ev{H} & 0 & y_1 \ev{\phi_1} & \tilde{y}_1 \ev{\phi_1} \\
          0 & y_1 \ev{\phi_1} &  y_2 \, \ev{\phi_2} & 0 \\
          0 &  \tilde{y}_1 \ev{\phi_1} &  0 & \tilde{y}_2 \,\ev{\phi_2} \\
        \end{pmatrix} \, .
      \end{equation}
At this point we emphasize the absence of mass terms for the singlet combinations $\bar{\nu}_R\nu_R^c$ and $\bar{N}_L N_R$ in the mass matrix \eqn \ref{eq:seesawInvDM} due to conformal invariance, see also \fig \ref{fig:CISS_Masses}. These terms would in principle be present in a non-conformal theory, and might be avoided by other extra discrete symmetries and new scalar particles. Note however, that many other phenomenological consequences to be discussed in this paper would not follow.

\noindent
Note that for simplicity we consider the one-flavour case, which can be straightforwardly generalized to the physical scenario with three flavours, then the Yukawa couplings will be $3\times 2$ and $2\times 2$ matrices, as we discuss shortly. The fermionic particle content comprises a left-handed Majorana fermion which is the active neutrino, a  pseudo-Dirac pair of right-handed neutrinos at the mass scale  set by $ y_1 \ev{\phi_1} =: M_R$ and a mass splitting of the order $  y_2 \, \ev{\phi_2} =:\m$, and a Majorana singlet neutrino of the mass $\m$. We will now demonstrate the diagonalization procedure of the above matrix which will lead to this mass pattern and discuss the mixing among the fermions.

Even though the $L \leftrightarrow R$ symmetry introduces relations among the Yukawa couplings and the induced masses it can be violated by higher-order interactions and for phenomenological purposes we consider the following induced mass matrix 
\begin{equation}
 \label{eq:seesawInvDM_Pheno}
        \mathcal{M} = 
        \begin{pmatrix}
          0 & m_D & 0 & 0 \\
          m_D & 0 & M_1 & M_2 \\
          0 & M_1 &  \mu_1 & 0 \\
          0 & M_2 &  0 & \mu_2 \\
        \end{pmatrix} \, ,
      \end{equation}
    with $M_1 \approx M_2$ and $\mu_1 \approx \mu_2$. To study the spectrum of this matrix we perform a rotation by an angle defined by $\tan \phi = \frac{M_1}{M_2}$ in the $\left(\nu_R^c, N_L, N_R^c   \right)$ subspace, which leads to the following structure
    
    \begin{equation}
 \label{eq:submatrix}
        \begin{pmatrix}
           0 & 0 & \sqrt{M_1^2+M_2^2} \\
           0 &  \frac{M_2^2 \mu_1 + M_1^2\mu_2}{M_1^2+M_2^2} & \frac{M_1 M_2 \left(\mu_1 - \mu_2\right)}{M_1^2+M_2^2} \\
           \sqrt{M_1^2+M_2^2} & \frac{M_1 M_2 \left(\mu_1 - \mu_2\right)}{M_1^2+M_2^2} & \frac{M_1^2 \mu_1 + M_2^2\mu_2}{M_1^2+M_2^2} \\
        \end{pmatrix} \, .
      \end{equation}
In analogy to the usual inverse seesaw scenario we observe that the heavy particle spectrum contains a pseudo-Dirac particle pair with mass of the order $ \sqrt{M_1^2+M_2^2} =:\mathbf{M}$    and a Majorana type mass splitting of the order $\frac{M_1^2 \mu_1 + M_2^2\mu_2}{M_1^2+M_2^2} =: \mu_D$. The other mass parameters we denote by $\frac{M_1^2 \mu_2 + M_2^2\mu_1}{M_1^2+M_2^2} = \mu_S$ and $ \frac{M_1 M_2 \left(\mu_1 - \mu_2\right)}{M_1^2+M_2^2}:=\delta M$. The difference to the usual inverse seesaw  is the existence of two states with Majorana masses. The mass matrix of these states is obtained by application of the usual seesaw formula under the assumption $ \lbrace m_D, \delta M \rbrace  \ll \mathbf{M} $ to the rearranged mass matrix $\left(\nu_L, N_L, \nu_R^c  , N_R^c \right)$
\begin{equation}
 \label{eq:lightseesaw}
        \mathcal{M} = \\
        \begin{pmatrix}
      0 & 0 & m_D & 0\\
      0 & \m_S & 0 & \delta M\\
      m_D & 0 & 0 & \mathbf{M}\\
      0 & \delta M & \mathbf{M} & \m_D\\
        \end{pmatrix} \, 
      \end{equation}
and yields the light neutrino mass matrix

   \begin{equation}
 \label{eq:lightseesaw}
        \mathcal{M}_{2\times 2} = \\
        \begin{pmatrix}
          \frac{ m_D^2  }{ \mathbf{M}^2}\,\mu_D & -\delta M \,\frac{ m_D  }{ \mathbf{M}} \\
        -\delta M \,\frac{ m_D  }{ \mathbf{M}} &  \m_S \\
        \end{pmatrix} \, .
      \end{equation}
The mass eigenvalues are 

\begin{align}
\label{eq:LightEigenvaluesFull}
& m_1 = \frac{1}{2} \left( \m_D  +\frac{\m_S m_D^2}{ \mathbf{M}^2}+ \right. \nonumber \\ 
& \left.-\sqrt{\m_D^2 -2 \m_D \frac{\m_S}{\mathbf{M}}m_D + 4 \frac{m_D}{\mathbf{M}^2} \delta M^2 + \m_S \frac{m_D^4}{\mathbf{M}^4}} \right),\\ \nonumber 
& m_2 = \frac{1}{2} \left( \m_D  + \m_S + \right. \nonumber \\ 
& \left. + \sqrt{\m_D^2 -2 \m_D \frac{\m_S}{\mathbf{M}}m_D + 4 \frac{m_D}{\mathbf{M}^2} \delta M^2 + \m_S \frac{m_D^4}{\mathbf{M}^4}} \right)
\end{align}
We assume that the exchange symmetry $L \leftrightarrow R$ in the Hidden Sector is broken by higher-order operators, however it is still approximatively present and allows to simplify the expressions by the use of $\mathbf{M} \approx M_1 \approx M_2$ and thus $\m_D \approx \m_S \approx \frac{\m_1+ \m_2}{2} =:\m_+$ and $\delta M \approx \frac{\m_1 - \m_2}{2} =:\bar{\m}$. Under this assumption we can expand in the small parameter $\frac{\bar{\m}}{\m_+}$ which leads to the eigenvalues 
\begin{align}
\label{eq:LightEigenvaluesApprox}
& m_1 = \m_+ \frac{m_D^2}{\mathbf{M}^2} - \frac{\bar{\m}^2}{\m_+} \frac{m_D^2}{\mathbf{M}^2}  + \frac{\bar{\m}^2}{\m_+}  \approx  \m_+ \frac{m_D^2}{\mathbf{M}^2} + \frac{\bar{\m}^2}{\m_+}\\ 
& m_2 =  \m_+ +  \frac{\bar{\m}^2}{\m_+}\frac{m_D^2}{\mathbf{M}^2}  - \frac{\bar{\m}^2}{\m_+} \approx   \m_+   - \frac{\bar{\m}^2}{\m_+}\,.
\end{align}
It is found that the active neutrino has a mass of the order $m_\text{active} \approx \mu \, \theta^2$, where $\theta$ is the active-sterile mixing and given by $\theta \approx \frac{m_D}{\mathbf{M}}$ with a perturbation of the order $ \frac{\bar{\m}^2}{\m_+}$.  The second state is a Majorana neutrino with mass at the $\mu$ scale and its mixing with the active neutrino is of the order $\tilde{\theta} \approx \theta \, \frac{\bar{\m}}{\m_+} $, and therefore additionally suppressed by the mass splitting induced by higher-order terms breaking the $L \leftrightarrow R $ symmetry in the Hidden Sector. We observe that, while the mass splitting among $M_1$ and $M_2$ has no dramatic effect on the physical observables the splitting $\mu_1-\mu_2 $ controls the coupling of the additional Majorana state at the $\mu$ mass scale and in the limit of exact $L \leftrightarrow R$ symmetry it even decouples. Thus for later phenomenological considerations it is reasonable to set $M_1 = M_2 = \mathbf{M}$. Since after symmetry breaking one of the Majorana degrees of freedom is responsible for the light neutrino mass, the mass splitting among $\mu_1$ and $\mu_2$ is of order of the light neutrino mass and thus in the eV range. This implies that the correction to the neutrino mass is of order $\frac{\bar{\mu}^2}{\mu_+} \approx 10^{-3} \text{ eV}$ which is within the experimental uncertainty. In addition it predicts an active sterile mixing of the keV mass state of $\tilde{\theta}^2 \approx \left( \frac{m_D}{\mathbf{M}} \frac{\bar{\mu}}{\mu_+} \right)^2 \approx 10^{-10} - 10^{-12}$, which we will compare to experimental constraints in \sect \ref{sec:LowEnergyParticlePheno}.

 The situation in the CISS is different from $B-L$ models \cite{Basso:2012ti,Humbert:2015yva} as the interactions $ \bar{N}_L\,\nu_R \,\phi_1$ and $\bar{N}_R^c\,\nu_R \,\phi_1^* $ violate lepton number explicitly. Lepton number is not a symmetry of the full theory, but turns out to be an accidental symmetry of the low energy SM sector.    
    
\section{\label{sec:RCSB} Radiative Conformal Symmetry Breaking and Implications} 

\noindent
The hidden sector is responsible for electro-weak symmetry breaking and the pseudo-Goldstone boson (PGB) associated with the conformal symmetry breaking has to reside mainly in the hidden sector, see for example \cite{Radovcic:2014rea}. In the case of one additional bosonic degree of freedom, the Higgs boson is mainly the PGB which phenomenologically requires larger values of quartic couplings and leads to low-scale Landau poles, see for example the discussion in \cite{Foot:2007ay}. This is not the case in the CISS. 
    
We will demonstrate the RCSB in our case. As discussed above the scalar field content is given by the $SU(2)$ doublet $H$ and two SM singlets $\phi_1$ and $\phi_2$. For simplicity we will use spherical coordinates in field space with the replacements
\begin{align}
\label{eqn:fieldDefs}
\phi_2 = r\, \sin \theta \sin \omega \, , \\ \nonumber
H = r \, \sin \theta \cos \omega\, ,  \\ \nonumber
\phi_1 = r \, \cos \theta\,.
\end{align}
We find with \eqn \ref{eqn:fieldDefs} and the definitions $(\tan \theta)^2 =: \epsilon$ and $(\tan \omega)^2 =:\delta$ that 
\begin{align}
& R(\Lambda): = (r\,\cos \theta \, \cos \omega )^4 \, V(r, \theta\, \phi_1) =    \\   \nonumber
&\frac{1}{2} \, \left((\delta+1)^2 \l_1  + \epsilon (2\,\delta (\delta +1 ) \l_{2\,1} +  2 (1+ \delta) \l_{H1} \right. \\
&\left.  + \epsilon (\delta^2 \l_2+ 2 \delta \,\l_{2H} +\l_H ) ) \right)\,. \nonumber  
\end{align}
The vanishing of this quantity at the scale of symmetry breaking $R(\Lambda_{RCSB})=0$ defines the classically flat direction in the potential, it is the renormalization condition. 
  
Assuming that the mixing among the scalars is not large i.e. $\epsilon, \, \delta < 1$ a hierarchical vev structure appears
\begin{align}
\ev{\phi_1} & = \ev{r}  (1+\epsilon)^{-1/2} = : v\,, \\ \nonumber
\ev{H} & = v \, \sqrt{\frac{\epsilon}{\epsilon+1}}\,, \\ \nonumber
\ev{\phi_2} &  = v\,\sqrt{\frac{\epsilon\,\delta}{\delta+1}}\,, \\ \nonumber
\Rightarrow  & \ev{\phi_1} > \ev{H} > \ev{\phi_2} \,.
\end{align}  
The scalar spectrum contains two massive excitations and one which is massless on tree level and corresponds to the flat direction in the potential. The idea behind the Gildener-Weinberg approach is that the quantum effects are taken into account in the one-loop correction to the mass of this particle, making it a PGB of broken conformal symmetry. This procedure ensures perturbativity as discussed in detail in \cite{Gildener:1976ih}. 

Expanding the fields about their expectation values we obtain the massive scalar spectrum, which has the following form on tree level
\begin{align*}
\frac{M_h^2}{ v^2 }   =  & \left(\epsilon  \left(\frac{3 \delta  \lambda _{12} \left(\lambda _H+5 \lambda
   _{12}\right)}{3 \lambda _H-\lambda _{12}}+  \right. \right. \\ \nonumber 
    & \left. \left. \frac{3 \lambda _{\text{H1}} \left(\lambda
   _H+5 \lambda _{\text{H1}}\right)}{3 \lambda _H-\lambda _{\text{H1}}}\right)+3 \lambda
   _H\right) \, \\ \nonumber 
\frac{ M_{\phi_2}^2}{v^2} =  &  \left(\epsilon  \left(-\frac{16 \lambda _{\text{H1}}^2}{3 \lambda _H-\lambda
   _{\text{H1}}}+3 \lambda _H+\delta  \lambda _{\text{H2}}\right)+\lambda
   _{\text{H1}}\right)\,.
\end{align*}
The PGB of the conformal symmetry breaking, which we will denote as Archaon from now on, acquires mass at the quantum level, which is parametrically suppressed
\begin{align}
\label{eqn:PGBMass}
M_{\phi_1}^2 = \frac{1}{8 \pi^2 \ev{r}^2}\left( M_{h}^4 + 6 m_W^4 + 3 m_Z^4 + 3 M_X^4 \right.\\ \nonumber 
\left. +M_{\phi_2}^4 -12 m_t^4 - 2 \sum_i M_{N_{i}}^4 \right) \,.
\end{align}
A possible configuration, which leads to the correct Higgs mass and the EW vev, has negative $\lambda_{H1}$ and $\lambda_{H2}$ and quartic couplings of the order $10^{-3}$. Therefore, the RG running remains stable and perturbatively treatable. An interesting observation is that at least one of the portal terms needs to be sizeable, of the order $\mathcal{O}\left( -0.1 \right)$, which makes the additional scalars accessible at the LHC. For the mass spectrum and the vevs we consider two benchmark points as numerical examples:
\begin{enumerate}
\item $\ev{\phi_1} = 1380 \text{ GeV}$, $\ev{H} = 246 \text{ GeV}$  , $\ev{\phi_2} = 38 \text{ GeV}$, $M_h = 125.5 \text{ GeV}$ and $ M_{\phi_2} = 2.17 \text{ TeV}$.
\item $\ev{\phi_1} = 1250 \text{ GeV}$, $\ev{H} = 246 \text{ GeV}$  , $\ev{\phi_2} = 181 \text{ GeV}$, $M_h = 124.9 \text{ GeV}$ and $ M_{\phi_2} = 3.06 \text{ TeV}$.
\end{enumerate}
The main differences among the scenarios are the vev hierarchies of $\ev{\phi_2}$ and $\ev{H}$. We find that in the allowed parameter region the vev of $\phi_2$ can be between $\mathcal{O} \left( 10\right)$ GeV and the electro-weak scale. Another observation is, that a large $\ev{\phi_1}$ leads to a heavy $M_{\phi_2}$.

\begin{figure}[h]
  \includegraphics[width=0.45\textwidth]{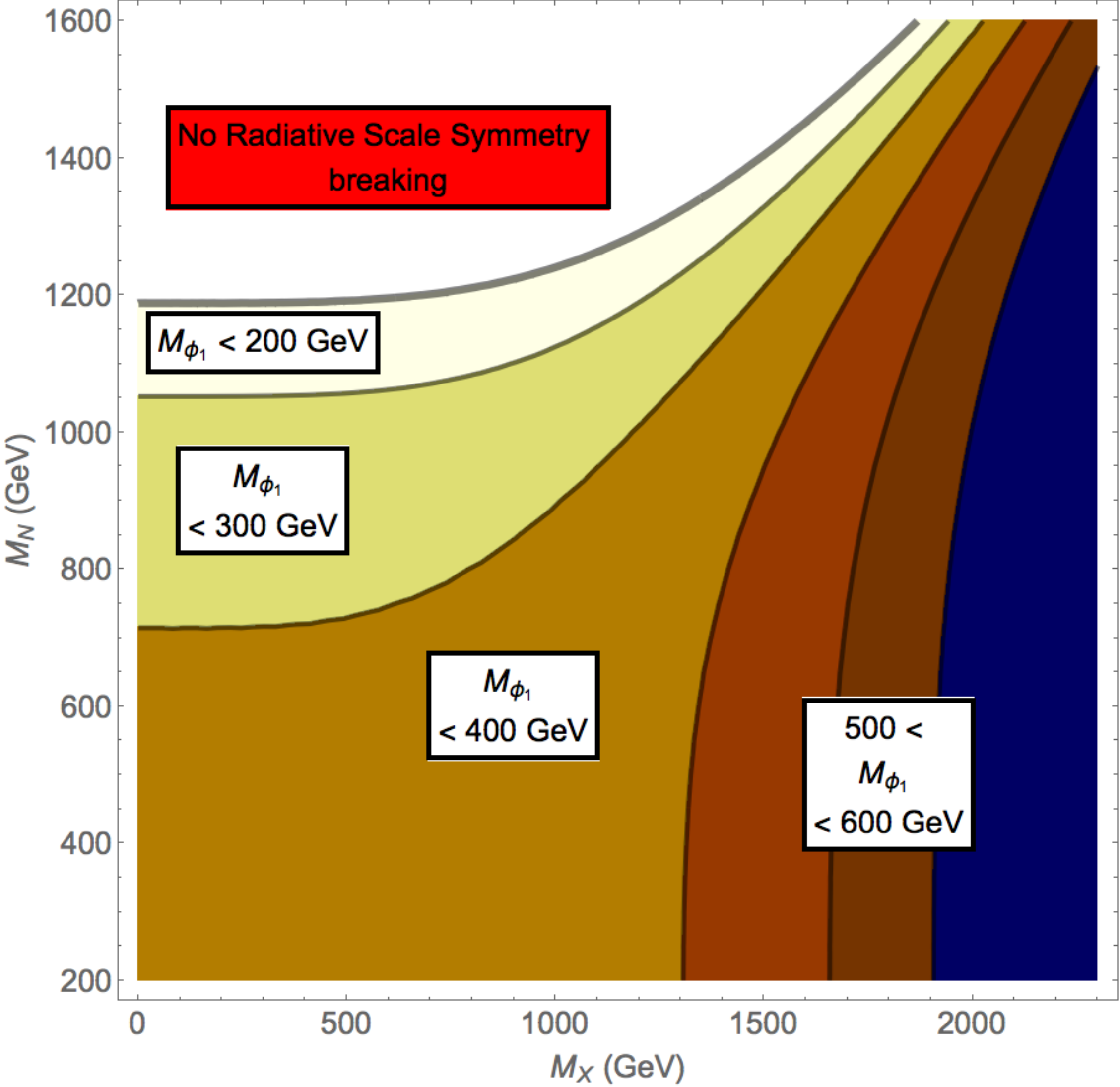}
  \caption{The phenomenlogically allowed mass region with RCSB, a Higgs mass of $125$ GeV, $M_{\phi_2} = 2 \text{ TeV}$, Higgs portal mixings compatible with the bound $\sin \theta < 0.37$, perturbative potential parameters and no low-scale Landau pole. Here $M_N$ is the average mass of the heavy right-handed neutrino, $M_X$ is the mass of the HS gauge boson and $M_{\phi_1}$ is the mass of the Archaon PGB. Note the upper bound on the right-handed scale of $1200$ GeV and the upper bound on the PGB mass of $400$ GeV for HS gauge boson masses below a TeV.} 
  \label{fig:ThreeMasses}
\end{figure}

In addition to the breaking of $SU(2)_L\times U(1)_Y$ symmetry the vev of $\phi_1$ breaks the $U(1)_X$ symmetry and generates a mass for the associated gauge boson $X$ of the order of the conformal symmetry breaking scale
\begin{align}
M_X =g_X  \sqrt{  \ev{\phi_1}^2 + 4 \ev{\phi_2}^2 } \approx g_X   \ev{\phi_1}  \,.
\end{align}
Considering the occurrence of spontaneous conformal symmetry breaking we find that for gauge boson masses below a TeV, there is an upper bound on the average mass for the heavy pseudo-Dirac neutrinos, which is $\bar{M}_N < 1200 \text{ GeV}$  and an upper bound on the induced Archaon PGB mass of $400$ GeV, see \fig \ref{fig:ThreeMasses} .  

\noindent
As can be seen from \eqn \ref{eq:TheModel} the Yukawa interactions of $\phi_1$ violate lepton number. If it was a global charge spontaneous breaking  would lead to a massless Goldstone boson with known consequences. As it is broken explicitly there is no potential problem with a massless Goldstone particle.

\section{\label{sec:Phenomenology}Phenomenology}

\subsection{Low energy Particle Physics Phenomenology}
\label{sec:LowEnergyParticlePheno}

\noindent
We consider three generations of active neutrinos and the minimal solution which can account for the oscillation phenomenology is a two-flavour set-up in the Hidden Sector, consistent with the findings in \cite{Abada:2014vea}. According to our discussion in \sect \ref{sec:Model} we then obtain a spectrum with two Majarona states at the intermediate scale $\mu$, and two heavy pseudo-Dirac particles at the mass scale \textbf{M} with a mass splitting of the order $\mu$. The latter account for two parametrically suppressed active masses, while the third active neutrino remains massless. In this section we identify regions in the parameter space allowed by low energy particle physics and comment on compatibility with astrophysical observations.

\subsubsection*{Active Neutrino Oscillations}
\noindent
To ensure that our morel is consistent with the oscillation phenomenology we use relation \eqn \ref{eq:LightEigenvaluesApprox}, and as we discussed earlier, the fact that $\bar{\m} < \sqrt{\m_+ \,10^{-3} \text{eV}}$. This leads to a parametrization of $m_D$, in a similar approach as in \cite{Casas:2001sr} 
\begin{align}
\label{eqn:Parametrization}
& \mathbf{1} = O(\theta)^T\,O(\theta) =  \nonumber \\ \nonumber
& m_\text{light}^{-1/2} U_\text{PMNS} \left( m_D^T \left( \mathbf{M}\, \m_+^{-1} \mathbf{M} \right)^{-1} m_D \right) U_\text{PMNS}^T \, m_\text{light}^{-1/2} \\ 
& \Rightarrow m_D = \mathbf{M} \, \sqrt{\m_+^{-1}}\, O(\theta) \sqrt{ m_\text{light} }\, U_\text{PMNS}^T\,,
\end{align}
where $m_\text{light}$ denotes the diagonal active neutrino mass matrix, $O(\theta)$ is a general $2\times 2$ orthogonal matrix and $\m_+$ can always be assumed diagonal with appropriate field definitions. An interesting observation is that even under the assumption that Yukawa couplings have a strong hierarchy, as in the charged lepton sector, neutrino masses can be much less hierarchical. In the conformal inverse seesaw all masses are generated due to Yukawa interactions and hierarchy in $m_D$ and in $\mathbf{M}$ can cancel, as can be seen from \eqn \ref{eqn:Parametrization} leading to reduced hierarchy among the light neutrino masses.

In addition to the oscillation phenomenology we require the following low energy constraints to hold. 

\subsubsection*{Non-unitarity}
\noindent
In the discussed model the active neutrino mixing matrix is no longer exactly unitary. This is a consequence of active-sterile mixing and it induces a number of effects on physical quantities as the Weinberg angle, the W-boson mass, the left- and right-handed couplings $g_L$, $g_R$, the leptonic and invisible Z-boson decay width and the neutrino oscillation probabilities, for more detailed discussion and limits see \cite{Antusch:2006vwa, Akhmedov:2013hec,Antusch:2014woa} and references therein. Thus studying the non-unitarity allows to narrow down the parameter space of a given model. For the study we define flavour dependent observables 

\begin{align}
\label{eq:NonUnitarityFlavour}
\e_\a = \sum_{i>4} |U_{\a i}|^2\,\,,\, \a \in \lbrace e,  \m, \tau \rbrace\,,
\end{align}
and the total non-unitarity measure $N := \e_\a + \e_\m + \e_\tau$.        
       
As given by \eqn \ref{eq:LightEigenvaluesApprox} the active-sterile mixing is determined by the ratio $m_D^2/\mathbf{M}^2$ and the general spirit of RCSB with a conformal symmetry breaking scale close to the EW scale suggests sizeable values. 
For ratios above $10^{-6}$ the phenomenology is considerably affected. The most sensitive observables are the Z boson invisible decay width and the Muon decay constant, which is used to determine the Fermi constant. The observables' dependence on the non-unitarity parameters (see \eqn \ref{eq:NonUnitarityFlavour}) is given by 
\begin{align}
\frac{\Gamma^\text{inv}_Z}{[\Gamma^{\text{inv}}_Z]_\text{SM}} = \frac{1}{3}\sum_{\alpha=e,\m,\tau} (1-\epsilon_\alpha)^2 \,,\\
G_\mu = G_F (1 - \epsilon_e)(1 - \epsilon_\mu)\,.
\end{align}
The region of sizeable active-sterile mixing with heavy particles is of particular interest, since contributions from heavy sterile neutrinos can improve the electro-weak fit, as discussed in \citep{Akhmedov:2013hec} and physical effects can be measurable. Requiring that the above observables are compatible with the experimental values, constraints for the model parameter space can be found. 

\subsubsection*{Lepton Universality}
\noindent
Various experiments, as discussed in \cite{Loinaz:2004qc}, show that the flavour dependent changes to the lepton couplings cannot differ too drastically. Thus we have effectively 
\begin{align}
& \e_e - \e_\m = 0.0022 \pm  0.0025 \,, \\ \nonumber
& \e_\m - \e_\tau = 0.0017 \pm  0.0038 \,, \\ \nonumber
& \e_e - \e_\tau = 0.0039 \pm  0.0040\,.
\end{align}
We demonstrate the impact of these constraints on the parameter space in \fig \ref{fig:LowEnergyAllowed}.

\subsubsection*{Lepton Number Violation and $0\nu \beta \beta$}
 
\noindent
In the CISS lepton number is violated explicitly by one unit in the interaction involving $\phi_1$. At the same time after symmetry breaking the same interactions of $\phi_1$ break $U(1)_X$ and violate $X$ by one unit. On the other hand the vev of $\phi_2$ violates $X$ by two units. The interaction among the fermions transfers this violation to the Lepton sector and thus lepton number is also broken by two units, making the  $0\nu \beta \beta$ decay possible. The lepton number violating decay $0\nu\beta\beta$, however, is in general suppressed in our scenario, as we will demonstrate. The general expression is \cite{Blennow:2010th} $\ev{m_{ee}} \approx |q^2 \textstyle{\sum_{i}} \mathbf{U}_{e i}^2 \, m_i/(q^2 - m_i^2)     |$ . Which now can be studied in three cases, depending on the ratio of $q^2/\mathbf{M}^2$, where the neutrino momentum is $|q| \approx  0.1 \,\text{GeV}$. 

If we have $\mathbf{M} \gg 0.1 \text{GeV}$ and using the facts that for  $i>5$, $\mathbf{U}_{e i}^2 \approx m_D^2/\mathbf{M}^2$  and $ \mu_+ \,m_D^2/\mathbf{M}^2  \approx m_\nu $ the following approximation holds, with $A_\text{PD} $ being the number of heavy  pseudo-Dirac states
\begin{align}
& \ev{m_{ee}}  \approx \left|\sum_{i=1}^3 \mathbf{U}_{e i}^2 \, m_i   + \sum_{i=4}^5 \frac{m_D^2}{\mathbf{M}^2}\frac{\bar{\mu}}{\m_+} \, m_i  - \frac{q^2}{2}  A_\text{PD} \mathbf{U}_{e\, 6/7}^2 \frac{\mu_+}{\mathbf{M}^2} \right| \nonumber \\ 
& \approx  \left|\sum_{i=1}^3 \mathbf{U}_{e i}^2 \, m_i   - m_\n \left( \frac{q^2}{\mathbf{M}^2} - \frac{\bar{\m}}{\m_+}\right) \right| \approx \ev{m_{ee}^\text{active}}\,,
\end{align}
which means that the rate is basically given by the light neutrino spectrum with well known phenomenology. 

The other limit is $\mathbf{M} \ll 0.1 \,\text{GeV}$, leading to $\ev{m_{ee}} \approx | \textstyle{\sum_{i}} (\mathbf{U}_{e i}^2 \, m_i  +  1/q^2\,\mathbf{U}_{e i}^2 \, m_i^3)    | = \mathcal{M}_{ee}+O(\mu_+\, m_D^2/q^2)$. Given that $\mu_+\, m_D^2/q^2 < \mu_+ \,\mathbf{M}^2/q^2$ the contribution of the additional states is negligible in this limit. 

The only case when the heavy pseudo-Dirac states can measurably contribute to the $0\nu\beta\beta$ decay is when $\mathbf{M} \approx  0.1\, \text{GeV}$. Then we have 

\begin{align}
\ev{m_{ee}} \approx \left|m_{ee}^\text{light} + \textstyle{\sum_{i>5}} \mathbf{U}_{e i}^2 \,\mu   \left( 1 + \frac{ m_i^2}{|q^2|} \right)^{-1} \right| \\ \nonumber
\approx   \left| m_{ee}^\text{light} + \textstyle{\sum_{i>5}}  m_\nu \, \left( 1 + \frac{ m_i^2}{|q^2|} \right)^{-1}\right|,
\end{align}
which is of the order of the light neutrino contributions. Thus we can understand why the limit by \cite{Gerda} of $\ev{m_{ee}}<0.4 \, \text{eV}$ only constraints light $\mathbf{M}$ masses of order GeV with considerable active-sterile mixing. Due to the cancellation in the pseudo-Dirac mass contribution this observable, however, does not severely constrain the parameter space of the CISS.

\subsubsection*{Lepton Flavour Violation and $\m \rightarrow e + \gamma $}

\begin{figure}[h]

  \includegraphics[width=0.45\textwidth]{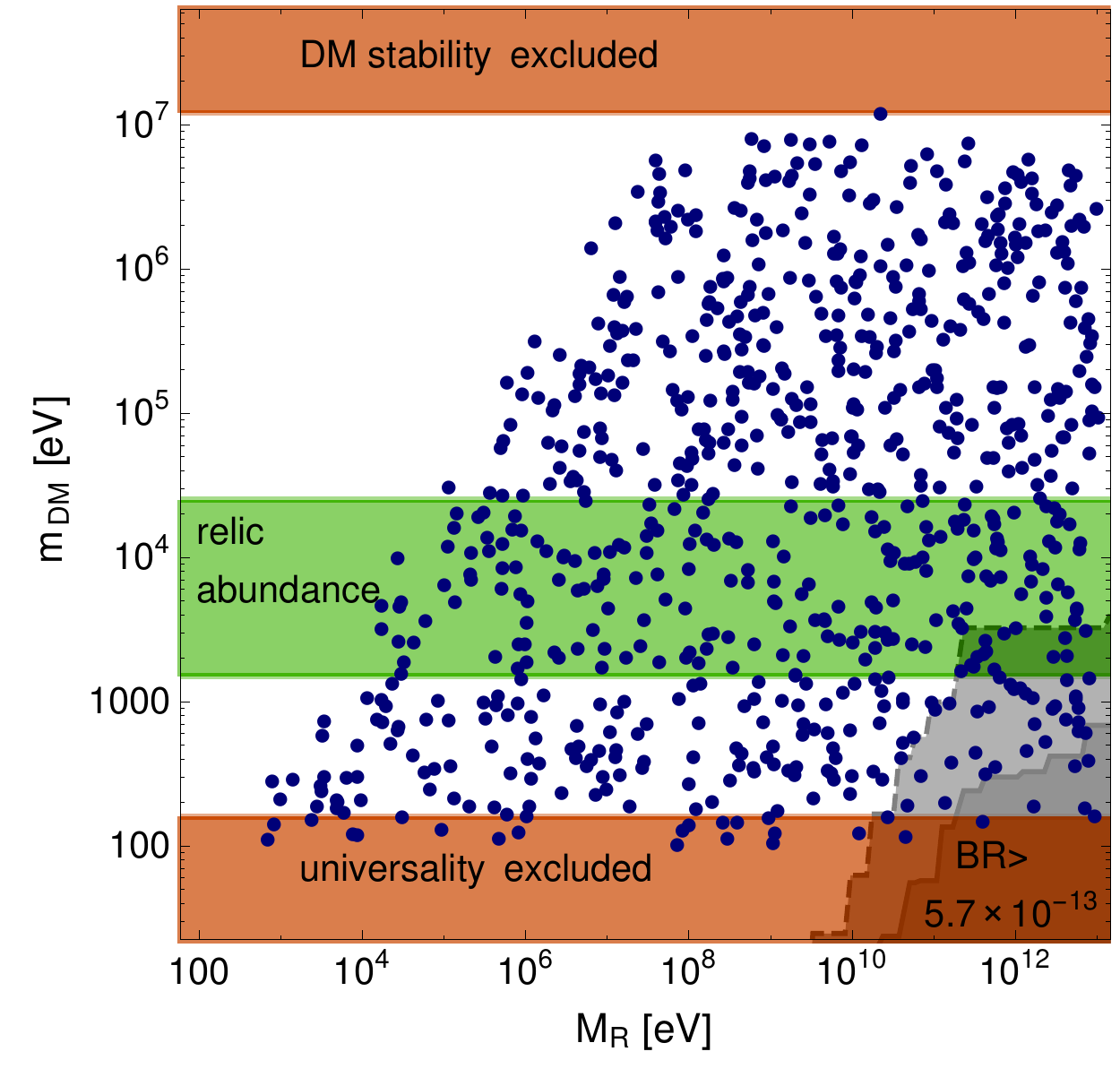}
  \caption{The heavy pseudo-Dirac mass scale (average mass) $M_R$ versus the DM mass scale $m_\text{DM}$. 
  The displayed points are from a scan showing cases allowed by low energy observables.
  The green region is singled out by the correct dark matter relic abundance from a non-thermal freeze-in mechanism. 
  Regions which are excluded by lepton universality and dark matter stability are displayed in brown.
  Furthermore, the grey-shaded area bounded by a solid (dashed) grey line shows the region where the branching ratio (BR) of $\mu \rightarrow e \gamma$ is excluded by the current (future) upper limit set by the MEG experiment for inverse light neutrino mass hierarchy.} 
  \label{fig:LowEnergyAllowed}
\end{figure}

\noindent
An interesting observation is that the suppression of lepton \textit{number} violating processes does not generically suppress lepton \textit{flavour} violating processes. The best constrained value is the branching ratio  $\text{Br}(\mu \rightarrow e + \gamma )$, where the limit is placed by the MEG collaboration \cite{Adam:2013mnn} and is $5,7 \cdot 10^{-13}$. The neutral fermion contribution to this loop-induced decay is
    
\begin{align}
\text{Br}(\mu \rightarrow e + \gamma ) =  \frac{3 \alpha_{\text{em}}}{32 \pi} \left| 2 \, \textstyle{\sum_{i}} \mathbf{U}_{\mu i}^* \mathbf{U}_{e i}
\,G \left(\frac{m_i^2}{M_W^2}\right)\right|^2,
\end{align}
with 

\begin{align}
& G(x) := \int_0^1 da \, ( 2 (1-a)(2-a) +  \nonumber \\ 
&   a (1+a)\,x )  (1-a )/((1-a) +x\,a)\,.
\end{align}
\noindent
Since in the loop function $G(x)$ the masses appear squared the cancellation leading to a suppressed $0\nu\beta\beta$ process cannot work.
We find that the MEG bound together with the lepton universality and neutrino oscillation constraints leads to the most severe limits on the model parameters, as shown in \fig \ref{fig:LowEnergyAllowed}. Note that the MEG collaboration has proposed an update of the experiment with a designated sensitivity of $\text{Br}(\mu \rightarrow e + \gamma ) < 6 \times 10^{-14}$ \cite{Baldini:2013ke}, which will lessen the available parameter space. 

Another flavour violating process is the decay $\mu \rightarrow 3e$. The current limit on its branching ratio is set by the SINDRUM collaboration and given by $\text{Br}(\mu \rightarrow 3e) < 1.0 \times 10^{-12}$ \cite{Bellgardt:1987du}. A new experiment, called ``Mu3e'' has been proposed with the aim to reach a sensitivity of $\text{Br}(\mu \rightarrow 3e) \sim 1 \times 10^{-16}$ \cite{Blondel:2013ia}. Note that in our model the branching ratio can be estimated by $\text{Br}(\mu \rightarrow 3e) \approx \text{Br}(\mu \rightarrow e + \gamma) \times \alpha_\text{em}$, since we do not have any particle leading directly to $\mu \rightarrow 3e$. This means that we expect $\text{Br}(\mu \rightarrow 3e)$ to be roughly by a factor of $100$ smaller than $\text{Br}(\mu \rightarrow e + \gamma)$. Comparing this to the future MEG sensibility we expect constraints of the same order of magnitude on the parameter space from $\mu \rightarrow 3e$.

\subsubsection*{Combined Limits}
\begin{figure}[h]

  \includegraphics[width=0.45\textwidth]{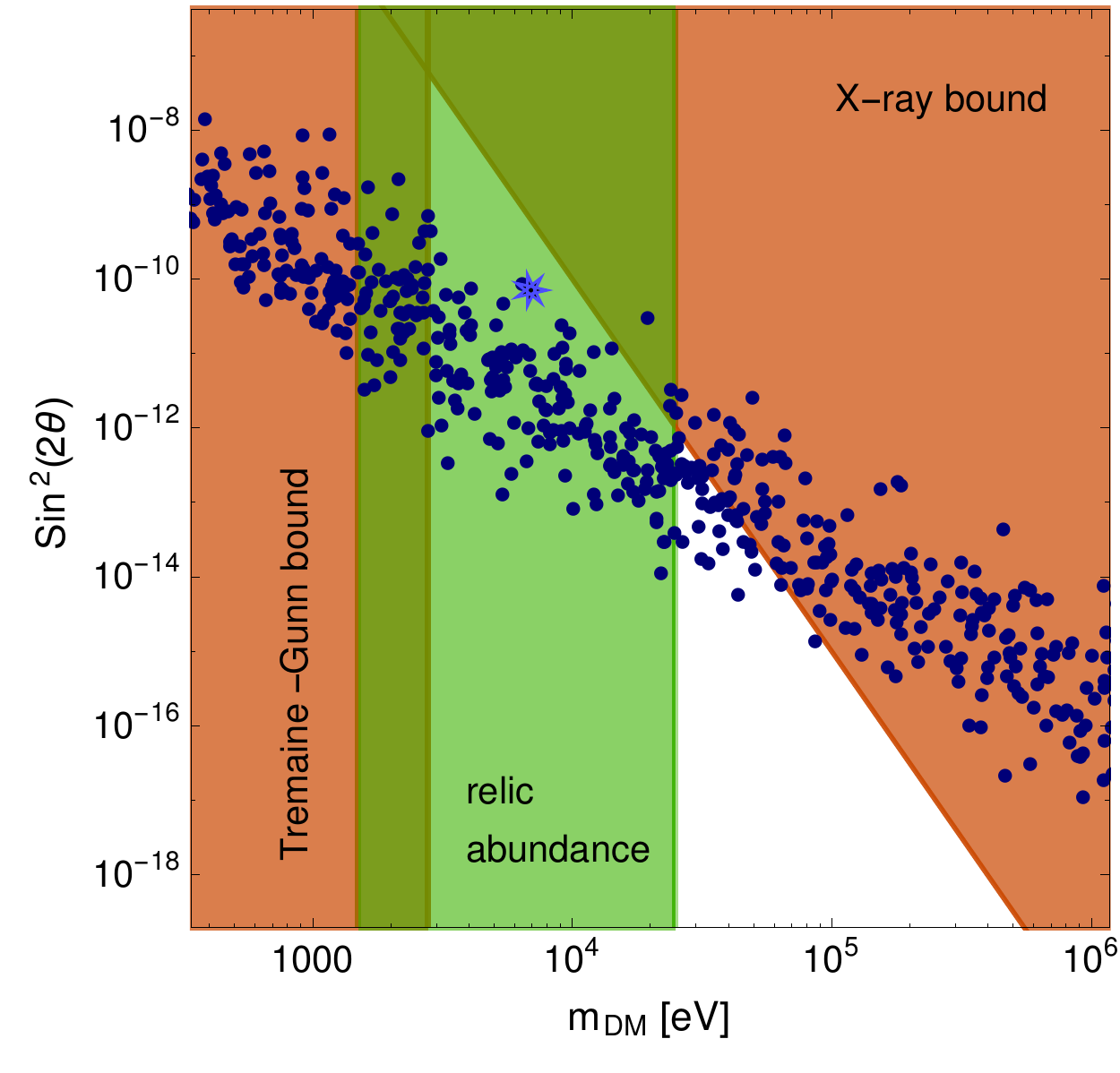}
  \caption{The dark matter scale $m_\text{DM}$ versus the mixing angle of keV sterile neutrino dark matter with the active neutrinos. 
  The region of parameter space of the CISS allowed by low energy observables is represented by the population of dots from a parameter scan. 
  The green region shows the mass scales compatible with a non-thermal freeze-in production mechanism discussed in \sect \ref{sec:RelicAbundance}, 
  which is the most generic production scenario in this model. Furthermore, the Tremaine-Gunn (TG) and the X-ray excluded regions are displayed in brown.
  The Lyman-$\a$ constraint turns out to be weaker than the TG bound in this case and is therefore omitted. 
  The claimed signal at 7 keV as discussed in \cite{Boyarsky:2014jta} is represented as a star.} 
  \label{fig:MasterPlot}
\end{figure}

\noindent
We observe that the combined limits from low energy particle physics with the requirement that the state at the intermediate scale $\m$ is produced in the early universe and is stable on cosmological scales i.e. $\tau_\text{Int.} > (10^2 ) \, \tau_\text{universe}$, leads to a window in the parameter space displayed in \fig \ref{fig:MasterPlot}. The astonishing observation is that this is exactly the region which is compatible with astrophysical requirements for a warm Dark Matter particle \cite{Abazajian:2006yn}, as the bound from X-ray observations, the phase space bound \cite{Tremaine:1979we}, the Lyman-$\a$ forest and several production mechanisms we will comment on in \sect \ref{sec:RelicAbundance}. Note that the allowed parameter region overlaps with the region where the Dodelson-Widrow mechanism \cite{Dodelson:1993je} does not produce hot DM, as discussed in \cite{Abada:2014zra}.

We present a mass spectrum of a benchmark point in the parameter space allowed by all phenomenological considerations. 

Pseudo-Dirac spectrum:
$M_{1/2} = 638 $ GeV with mass splitting of 10 keV and  $M_{3/4} = 9.25$ GeV with mass splitting of 9 keV.

Intermediate scale spectrum: 
$M_5 = 7.013$ keV and  $M_6 = 7.006$ keV 
with active-sterile mixing $\sin^2{(2\theta_5)} \approx 7 \cdot \,10^{-11}$ and $\sin^2{(2\theta_6)} \approx 3.2 \cdot \,10^{-13}$.

Light active spectrum:
$M_7 = 0.049$ eV, $M_8 = 0.0085$ eV and  $M_9 \approx 0$.

The non-unitarity is $\epsilon \approx 10^{-5}$, the effective mass for $0\nu \b \b$ is $\ev{m_{ee}} \approx 0.003$ eV and the branching ratio $\text{Br}(\mu\rightarrow e \gamma) \approx 1.01 \cdot 10^{-13}$. 

We find that from the low energy particle physics perspective the most accessible observable seems to be the branching ratio of $\mu \rightarrow e \gamma$.

\subsection{Collider Phenomenology of the Hidden Sector}

\subsubsection*{Pseudo-Dirac Neutrinos}
\noindent
The most promising signature to distinguish the heavy pseudo-Dirac neutrino of the CISS scenario from a heavy Majorana neutrino is a direct test at a collider, which is feasible as all the fermions involved are below the TeV scale. The difference lies in the dominant decay channel of the right-handed neutrinos. Since in the Type-I Majorana seesaw the lepton number violation is unsuppressed, the dominant process is expected to be the lepton number violating decay, see \cite{Keung:1983uu} and \cite{Abada:2013aba,Abada:2014cca, Antusch:2015mia}. As argued in \cite{Almeida:2000pz} the relevant quantity to estimate the efficiency of the LHC concerning the Majorana neutrino detection is $\left| \sum_{i\, \text{hevy}} \mathbf{U}_{e i}^2 \frac{1}{M_i} \right| \geq 6\cdot 10^{-3} \text{ TeV}^{-1} $. In the CISS case there are two heavy pseudo-Dirac neutrinos and the sum simplifies to $\left|  \sum_{i = 1,2 } \mathbf{U}_{e i}^2 \left(\frac{1}{M_i}-\frac{1}{M_i+\m} \right)\right| \approx \left| \mathbf{U}_{e 1}^2 \frac{\m}{M_1^2} +\mathbf{U}_{e 2}^2 \frac{\m}{M_2^2} \right| \approx \epsilon \frac{\m}{\mathbf{M}^2 } $ due to the cancellation among the masses. Since for the process to be relevant $M_i > M_W$ and $\m$ is at the keV scale, the suppression factor of $\frac{\m}{\mathbf{M} } \leq 10^{-8}$ makes the signal irrelevant for phenomenology \cite{Kersten:2007vk}. 

\begin{figure}[h]

  \includegraphics[width=0.45\textwidth]{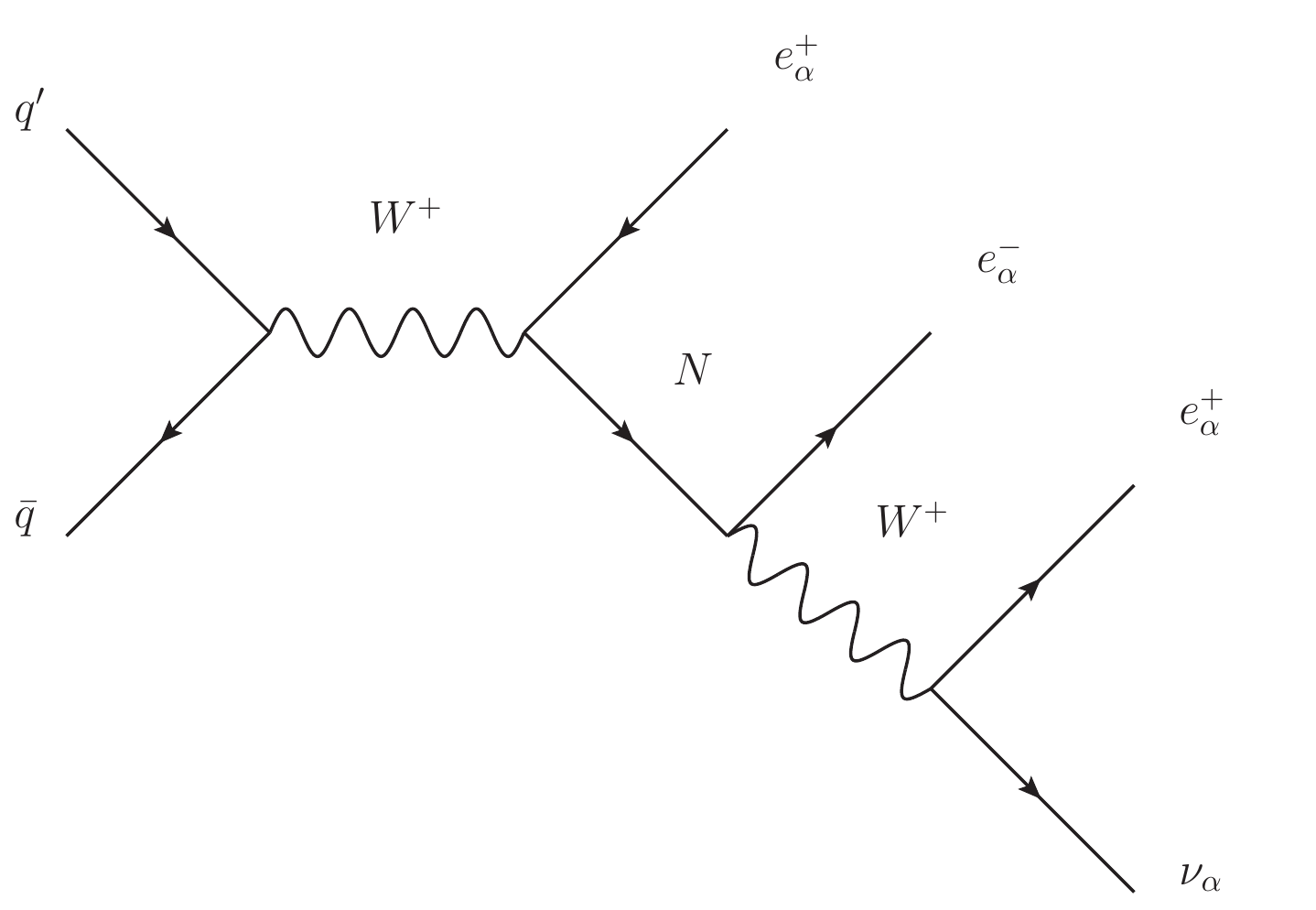}
  \caption{New collider signature for the ISS scenario with the trilepton plus missing energy signature.} 
  \label{fig:DecayDirac}
\end{figure}

As argued in  \cite{Das:2012ze, Das:2014jxa}, the most interesting channel to consider in case of suppressed same sign dilepton signal is the trilepton decay with missing energy, see \fig \ref{fig:DecayDirac}, since its SM background is significantly lower.
As can be seen from the Feynman graph in \fig \ref{fig:DecayDirac}, this decay also crucially depends on the active-sterile mixing squared and thus on the non-unitarity parameter $\epsilon_\alpha$. The interesting feature of the CISS in the RCSB framework is, that a large-scale separation is not expected which results naturally in an active-sterile $\epsilon \approx \theta^2 \approx m_D^2/\mathbf{M}^2$. Thus the most natural value for $\epsilon$, given an order of magnitude between the scales and Yukawa couplings of order one, can be up to one percent. The sizeable active-sterile mixing is compatible with the excess observed in the dilepton channel \cite{CMS:2014jfa}, as we will show in the next subsection, and thus a similar excess is expected to appear in the trilepton decay. Note that the excess of $30 \pm 10$ events as reported in \cite{Khachatryan:2014dka} could be accounted for by a Feynman graph similar to \fig \ref{fig:DecayDirac}, but with the secondary W boson decaying into jets. The number of events produced by this interaction is expected to be small due to the off-shell W boson. 
Furthermore, due to suppressed lepton number violation it is clear within the CISS why no corresponding excess in the same-sign dilepton channel has been observed \cite{Chatrchyan:2013fea}, as is expected in the case of a decaying $W_R$. If this pattern was confirmed in the next run of the LHC, it would be a strong point in favour of the CISS.

Note that the recently proposed production mechanism for heavy sterile neutrinos via t-channel processes can further increase the collider sensitivity and test mixings of $\epsilon_e \approx 10^{-4}$ for masses in the few hundred GeV regime, as argued in \cite{Dev:2013wba}.

\subsubsection*{Scalars}
\begin{figure}[h]

  \includegraphics[width=0.45\textwidth]{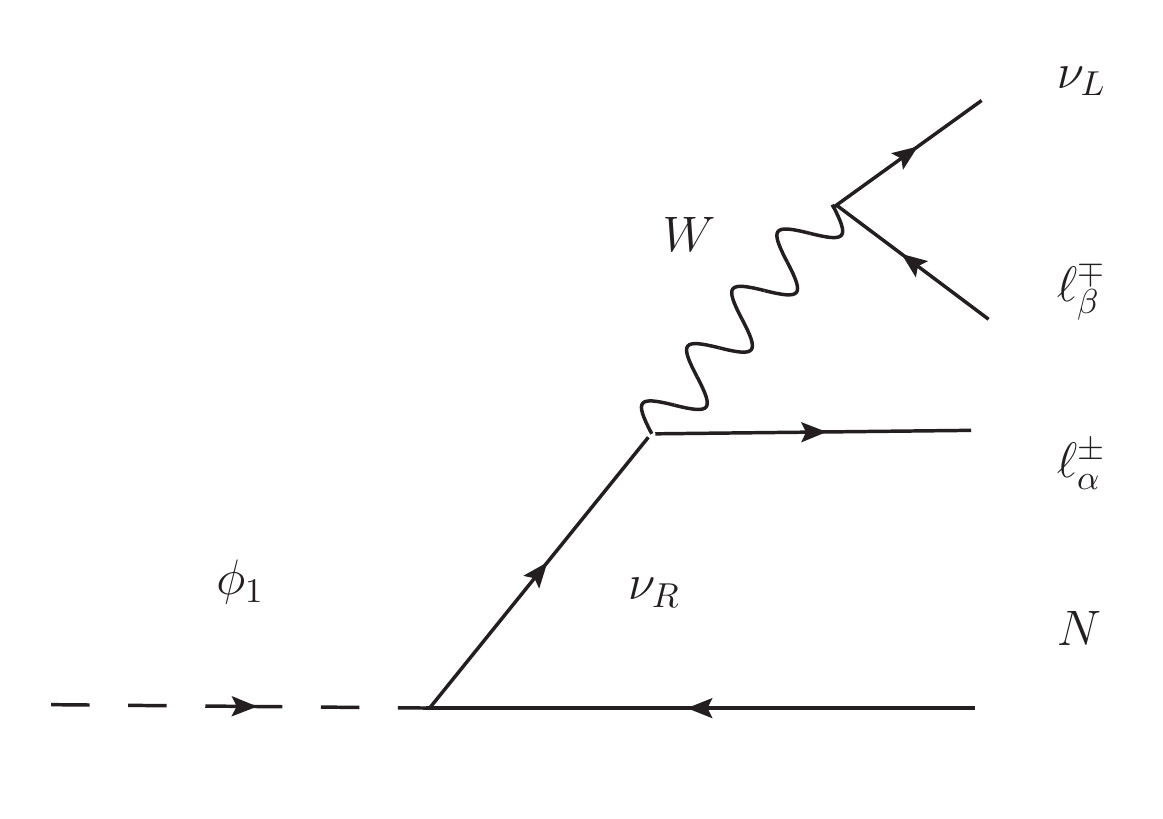}
  \caption{The new decay channel of the Archaon scalar  $\phi_1$. This digram shows a decay of the scalar partially to the visible sector, while the W decays leptonically. However, all other combinations of decays to two DM particles or to two leptons with two Ws are possible as well.} 
  \label{fig:HigghsAndLNV}
\end{figure}

\noindent
We found that in a generic situation one scalar has a mass in the few TeV region, while the mass of the Archaon PGB can be as light as a few hundred GeV. For example we found, that for the HS gauge boson masses below a TeV the PGB has to be lighter than 400 GeV. Additionally for the conformal symmetry breaking to be transmitted to the EW sector there has to be at least one portal coupling which is of the order $\mathcal{O}\left( 0.1\right)$ which means that the PGB could manifest itself at the LHC as a second Higgs like particle with the signal strength reduced roughly by a factor of hundred. Due to the heavier mass the signal is most likely to manifest itself in the $t \bar{t}$ system. As the Yukawa coupling $y_1$ is of order unity, $\phi_1$ will also have decays via HS particles, see for example  \fig \ref{fig:HigghsAndLNV}. The second scalar $\phi_2$, on the contrary, has very small Yukawa coupling $y_2\approx \mathcal{O} (10^{-7})$ to the HS and thus will mainly decay through the Higgs portal.  

The possible HS decay channels of the Archaon are $\phi_1 \rightarrow N_{DM}\nu_R \rightarrow a \ell^\pm + \text{Jet(s)} + \slashed{E}_T$, where $a \in [0,4]$ is the number of produced charged leptons. The jet multiplicity in the $\phi_1$ decays is not fixed due to initial state radiation independent of the respective decay. In all leptonic decays of $\phi_1$, except for the decay $\phi_1 \rightarrow 2 \ell^\pm + 2 (q^\prime \bar{q})^\mp$, final state neutrinos are produced leading to missing transverse energy $\slashed{E}_T$. The estimated decay rates are
\begin{align*}
&\Gamma_\text{Top} = \frac{3\,y_t^2 M_{\phi_1} \l_p^2}{ 8 \pi} \sqrt{1 - \frac{4 \,m_t^2}{M_{\phi_1}^2}} \, ,\\ 
&\Gamma_{\alpha\,\beta}\approx \frac{y_1^2 M_{\phi_1} s f^2 \left| \sum_{i, \text{heavy}}\mathbf{U}^*_{\alpha i}  \mathbf{U}_{\beta i} \right|^2}{8 \pi} \sqrt{1 - \frac{4 \,M_i^2}{M_{\phi_1}^2}}\,,
\end{align*}
where in the second line $s$ denotes a symmetry factor and $f$ accounts for the relative strength of the corresponding decay channel. Regarding the tentative measurement at the LHC of an excess in the decay channel $e^\pm e^\mp  + \text{Jets} + \cancel{\it{E}}_{T} $ \cite{CMS:2014jfa}  of $(130 \pm 50)$ events at $2.6$ standard deviations,  we can estimate the parameter values in the CISS to account for this observation. We find that the production cross section for the Archaon $\phi_1$ should be about $2.5\cdot10^{-1} \text{ pb}$. Assuming that the production is analogous to the Higgs boson, but suppressed by the portal coupling, we estimate that for $M_{\phi_1}$ in the 500 GeV region one needs $\lambda_p^2 \approx 0.25$, if we take the expected Higgs cross section as in \cite{Anastasiou:2012hx}. The mixing matrix elements of the heavy states to the active neutrinos is required to be of about $\epsilon_e\approx \mathcal{O}(10^{-2})$ and to non-active neutrinos $\mathbf{U}_{N i}^2\approx 0.8-1.0$, which is in agreement with the DM phenomenology. As an example we take the benchmark scenario of $\epsilon_e \approx 0.017$ and $U_{Ni}^2 \approx 0.97$ , for which the relevant branching ratios are $\text{BR}(\phi_1 \rightarrow 2 \ell +\text{Jet(s)} + \slashed{E}_T) \approx 1.5~\%$ and $\text{BR}(\phi_1 \rightarrow t\bar{t}) \approx 2.1~\%$ producing a signal of about 75 events in the $\phi_1 \rightarrow 2 \ell +\text{Jet(s)} + \slashed{E}_T$ channel. Due to the small BR($\mu \rightarrow e + \gamma$) we do not expect any direct decays into muons, but that the produced leptons are mainly electrons. Note, however, that a small fraction of $\tau 's$ can well be produced, which themselves decay into $e's$ and $\mu 's$ each with a branching ratio of roughly $20~\%$ \cite{Agashe:2014kda}.
We observe that the parameters needed to explain the measured excess would also lead to an excess of about 100 events in the $t \bar{t}$ decays in the 500 to 600 GeV region, which is in agreement with current uncertainties \cite{Chatrchyan:2013lca,ATLAS-CONF-2013-099}. 
 
In the next LHC run the model hypothesis should manifest itself in the $t \bar{t}$ system as a signal with 500 to 600 GeV invariant mass. At a designated integrated luminosity of $\sim 100\text{ fb}^{-1}$ $(\sim 300\text{ fb}^{-1})$ in the year 2018~(2021) \cite{Chatrchyan:2008aa,CMS:2013xfa}, we predict a signal of $390~(1160)$ events in the $\phi_1 \rightarrow 2 \ell +\text{Jet(s)} + \slashed{E}_T$ channel and a signal of $520~(1560)$ $t\bar{t}$ events using the branching ratios given above.

A different test for the size of the Higgs portal coupling can be performed in a general way by considering the Higgs couplings to the SM particles. The effective Lagrangian reads 
\begin{align}
&\mathcal{L}_\text{eff} = (1 + \rho )\,C_{H\,W\,W} H\,W_\mu \,W^\mu + (1+ \rho) C_{H\,Z\,Z} H\,Z_\mu Z^\mu  \nonumber \\ 
&- (1+\rho ) C_{H\,b\,b} H\,\bar{b} b  - (1+\rho ) C_{H\,\tau\,\tau} H\,\bar{\tau} \tau \nonumber  \\ 
&+ (1+\rho) C_{H\,g\,g} H \,G_{\mu\nu}G^{\mu\nu} + (1+ \rho)C_{H\,\gamma\,\gamma} H A_{_\mu\nu} A^{\mu\nu}  \nonumber  \\
& - (1+\rho ) C_{H\,c\,c} H\,\bar{c} c -\rho \, C_{H\,t\,t} H\,\bar{t} t\,.
\end{align}
The coefficient $\rho \approx -\frac{1}{2} \theta^2$ with $\theta$ the sum of mixing angles of the Higgs to additional scalars, is a universal suppression factor. A global fit to the data can lead to a bound on the mixing parameter, which is currently $\sin \theta < 0.36$ \cite{Farzinnia:2013pga}. 

\subsubsection*{Hidden Sector Gauge Boson}
\noindent
The gauge boson associated with the breaking of the Hidden Sector $U(1)_X$ symmetry can be searched for mainly in processes where it is produced due to the $U(1)$ mixing controlled by the parameter $\kappa$ in \eqn \ref{eq:TheModel}. As the conformal symmetry breaking scale sets also the scale of the $U(1)_X$ breaking, the new boson is expected to have a TeV scale mass, and thus to be well within reach of the LHC. The HS gauge boson possesses the same decay channels as the $\phi_1$ except for the $t\bar{t}$ channel
. It is, however, not produced over the Higgs portal, but from the mixing of the $U(1)_X$ gauge boson with the SM $U(1)_Y$ gauge boson leading to a different production cross section.

The $X$ decay channels with the most visible events are the $1 \ell$ and $2 \ell$ channels. If we take the parameters as given for the $\phi_1$ decays above to explain the excess of \cite{CMS:2014jfa}, we find an upper limit of $\kappa < 0.02$ in order to be consistent with \cite{ATLAS-CONF-2014-017}.

Concerning the boson masses of the HS we analyse \eqn \ref{eqn:PGBMass} with the mass pattern of $M_{\phi_1} \approx 550 \text{ GeV}$ and $M_{N_i} \lesssim M_{\phi_1}$ to account for the $\phi_1$ decay signal. We find a lower bound of $M_X \gtrsim 1 \text{ TeV}$ for values of $M_{\phi_2} \lesssim 2 \text{ TeV}$.

\subsection{Dark Matter Relic Abundance}
\label{sec:RelicAbundance}
\noindent
In this section we will discuss how the Dark Matter relic abundance in the CISS can be generated to explain the required abundance of Dark Matter. 
Depending on the details of the realization there are three possible mechanisms, which could account for the correct relic abundance. 

\begin{enumerate}
\item Production through oscillations in the early plasma, known as the Dodelson-Widrow~(DW) mechanism \cite{Dodelson:1993je}. The generic realisation of this mechanism, however, cannot account for the full amount of DM, as it is already severely constrained from structure formation observations. The possibility of a  resonant production with a large Lepton asymmetry in the early universe is still allowed by data \cite{Asaka:2005an}. This scenario requires adjustment in the parameters, which is of course not excluded a priori but can make it less attractive from the theoretical perspective.

\item As the Hidden Sector has a gauged symmetry broken by the scalar vevs there is a new massive vector boson, which can thermalize the DM candidate and if the gauge boson mass is sufficiently low the DM will be overproduced. The subsequent injection of entropy by the decay of TeV scale right-handed neutrinos, which can be heavy pseudo-Dirac states in the CISS, allows to avoid the overclosure of the universe, as discussed in \cite{Bezrukov:2009th,Nemevsek:2012cd,Abada:2014zra}. This mechanism also requires a conspiracy between model parameters and is thus not a generic feature.

\item At last we would like to point out that a generic mechanism in the CISS framework does exist, namely the non-thermal freeze-in production.  We will now discuss this mechanism in more detail.
 
\end{enumerate}

It turns out that in the CISS the relic abundance of warm Dark Matter is achieved naturally through a freeze-in mechanism. The first observation is that the scale of the vev which generates the intermediate keV scale is between the GeV and EW scales, which means that the Yukawa coupling is of the order $10^{-7}\, - \, 10^{-8}$. Thus this coupling never thermalizes the keV scale state in the early universe. This also implies that the effective number of relativistic degrees of freedom is unaffected.

The keV state is produced in a decay of the scalar, which generates its mass. The production through the decay of the scalar particle dominates over the decays of the sterile neutrinos, as shown by \cite{Chu:2011be}. The construction of our potential in \eqn \ref{eqn:fieldDefs} is such that the vevs are hierarchical, beginning at the TeV, and going to the EW scale. The smallest vev is between 10 and 200 GeV and due to the hierarchy the mixing among the scalars is in the $0.1$ region. Therefore, in comparison to the mechanism proposed in \cite{Kang:2014cia} no cancellation among the scalars can occur. This cancellation, however, is not necessary, as the decaying scalar has a vev below the EW scale. We find that the relic density can be calculated as discussed in \cite{Hall:2009bx,Merle:2013wta,Klasen:2013ypa} as 

\begin{align}
Y_X (\infty) \approx \frac{45 \, g_\text{int}}{1.66 \pi^4 g^{S}_{*} \sqrt{g^\rho }} \frac{\Gamma(7/2)\,\Gamma(5/2)\, M_{Pl}}{16\,M_\phi^2} \, \Gamma\left( \phi \rightarrow  N\, N \right)\,,
\end{align}
and leads in our scenario to the following simple relation
\begin{align}
\Omega_{DM\,h^2} \approx 0.11 \, \left(\frac{m_{DM}}{10\,\text{keV} }\right)^3 \left(\frac{\text{TeV}}{\ev{\phi_2}}\right)^2 \left( \frac{100 \text{GeV}}{M_{\phi_2}}\right)  \frac{10^3}{g^{S}_{*} \sqrt{g^\rho }} \,,
\end{align} 
where $g^{S}_{*}, g^\rho$ are the number of degrees of freedom active at $T \approx M_\phi$ relevant for the entropy and energy density. 

We can deduce limits on the Dark Matter particle mass from the requirement that the freeze-in leads to a relic density compatible with observations \footnote{Note that the DW mechanism will lead in our parameter regime to a production of approximately one third of the relic density, as discussed in \cite{Abada:2014vea}. Nevertheless, our considerations are valid to estimate the approximate mass required for the Dark Matter particle.}. Given that in the CISS the SM is augmented by 16 additional degrees of freedom we find that $10^3 / \left( g^{S}_{*} \sqrt{g^\rho } \right) \approx 1 $. As we have observed that the vev of $\phi_2$ is between 10 and 200 GeV with the mass $M_{\phi_2}$ in the few TeV regime we find that the Dark Matter mass has to be $1.5 \,\text{keV} < m_{DM} < 25 \, \text{keV}$.

Two comments are in order. Firstly, for the discussed mechanism to be at work the DM particle must not be thermalized by the HS gauge interactions, which means that the combination of the gauge coupling over the gauge boson mass has to be sufficiently small i.e., as $g_X/M_X \approx \ev{\phi_1}^{-1}$, $\ev{\phi_1}$ has to be above the TeV scale. 

Secondly, as discussed in our model there are two states with keV scale masses. We found that there is a hierarchy in the active-sterile mixing of the keV states and the active neutrinos. Therefore, one of the states will not be produced in the DW mechanism and will thus be less abundant by at least 30 $\%$. For astrophysical observations this means that a line signal from the DM decay will lead to a slightly asymmetric double line, with a sub-keV energy splitting.

\section{\label{sec:Conclusion}Conclusion}

\noindent
Motivated by the current experimental situation different realizations of conformal electro-weak 
symmetry breaking have been discussed recently by various authors. In the conformal framework there would be 
important consequences for the neutrino sector, since no explicit Dirac or Majorana mass term 
would be allowed in the Lagrangian. All Dirac and Majorana mass terms had to arise then from
Yukawa couplings times vacuum expectation values of suitable scalars. We presented a simple 
extension of the SM which realizes in this framework the so-called inverse seesaw mechanism. 
This model can nicely account for non-zero neutrino masses and spontaneous conformal symmetry 
breaking while avoiding  Landau poles in running couplings. 

The discussion of radiative conformal symmetry breaking via a portal to some hidden sector leads to 
scenarios where the driving scalar scales are in un-tuned cases generically in the multi-TeV range. 
The portal communicates this scale then to the visible sector which sets the electro-weak vacuum 
expectation value. 
We discussed in this paper the Conformal Inverse Seesaw (CISS), which is a very natural model 
for the explanation of small neutrino masses without extremely tiny Yukawa  couplings . 
In this scenario the explicit lepton number violation in the Hidden Sector (HS) is cast down to the active neutrino sector by spontaneous symmetry breaking of the U(1)$_X$ gauge group through the vacuum expectation values of the HS scalars. It is controlled by Yukawa interactions with a small coupling constant ($y_2 \sim 10^{-7}$), which is natural in the t'Hooft sense. It is a remarkable feature of the model that, since lepton number is not a conserved quantum number in the first place, there is no lepton number violation scale (as would be the case in a theory with e$.$g$.$ a broken U(1)$_{B-L}$), but instead lepton number violating processes are suppressed by the seesaw relation of the CISS. At the same time the inverse seesaw at the TeV scale naturally leads to a long-lived Dark Matter particle at the keV scale, which is consistent with the warm Dark Matter scenario. 

The spectrum of the model comprises of two pseudo-Dirac neutrinos of the scale $\mathbf{M}$, which naturally is at the TeV scale. The light neutrino mass is given by \eqn \ref{eq:LightEigenvaluesApprox} and the additional two sterile states have a mass of  $\mu \approx \text{keV}$ and a small mixing with the active neutrinos suppressed by $\Delta \m$, which vanishes in the limit of exact $L \leftrightarrow R$ exchange symmetry in the Hidden Sector. The dominant interaction in that case is the Yukawa coupling to the scalar, which generates the mass for the $\mu$ scale state. 

The remarkable feature is that the scale $\mu \approx \text{keV}$ required by the seesaw relation is also the correct scale for this state to be a Dark Matter candidate \cite{Dolgov:2000ew, Bezrukov:2009th}. Furthermore, the parameter region allowed by low energy observables and non-thermal production overlays exactly the region allowed by astrophysical experiments and the phase space density considerations, as discussed in \sect \ref{sec:LowEnergyParticlePheno}.   

We find that the CISS can be tested in low energy particle experiments, with the $\mu \rightarrow e \gamma$ measurement being the most promising experiment in the near future. Furthermore, we argue that the pseudo-Dirac states can be produced at the collider if the active sterile mixing is sizeable and their mass is above the W-boson mass. At the same time the decays of the Archaon $\phi_1$ and the HS gauge boson X, may already have been detected at the LHC leading to the excesses in the di-electron final states.   

Concluding we find that incorporating neutrino mass generation in radiative conformal symmetry breaking leads to very interesting and testable consequences.

\section*{Acknowledgments}
\noindent
We would like to thank Alexei Smirnov, Werner Rodejohan and Rhorry Gauld for helpful discussions.



\end{document}